\documentclass[iop]{emulateapj}
\shorttitle{Hoyle-Lyttleton Accretion in 3D}
\shortauthors{Blondin \& Raymer}

\begin{document}

\title{Hoyle-Lyttleton Accretion in Three Dimensions}

\author{John M. Blondin and Eric Raymer}
\affil{Department of Physics, North Carolina State University, Raleigh, NC 27695-8202, USA}
\email{John\_Blondin@ncsu.edu}

\keywords{accretion---hydrodynamics---shock waves}


\begin{abstract}

We investigate the stability of gravitational accretion of an ideal gas onto a compact object moving through a uniform medium at Mach 3.
Previous three-dimensional simulations have shown that such accretion is not stable, and that strong rotational 'disk-like' flows
are generated and accreted on short time scales.  We re-address this problem using overset spherical grids that
provide a factor of seven improvement in spatial resolution over previous simulations. 
With our higher spatial resolution we found these 3D accretion flows remained remarkably axisymmetric.  
We examined two cases of accretion with different sized accretors.
The larger accretor produced very steady flow, with the mass accretion rate varying by less than 0.02\% over 30 flow times.  The smaller accretor exhibited an axisymmetric breathing mode that modulated the mass accretion rate by a constant 20\%.  Nonetheless, the flow remained highly axisymmetric with only negligible accretion of angular momentum in both cases.

\end{abstract}

\section{Introduction}

The gravitational accretion of an ideal gas onto a compact object has been applied across a wide range of astrophysical topics, 
including the growth of supermassive black holes and associated feedback in AGNs \citep{booth}, accretion onto 
young star clusters as an explanation for the formation of ultraluminous x-ray sources \citep{nrrl09} and 
accretion onto gas giant planets around evolved stars \citep{villaver}. 
A common application is that of wind-fed X-ray binaries, beginning with the original suggestion by \citet{do73}
that the X-ray luminosity of Cyg X-3 is driven by the gravitational capture of material in the stellar wind of the companion star. 
More recently, wind-fed accretion onto neutron stars has been invoked to explain the behavior of supergiant fast x-ray transients \citep{ducci09}.

While the basic theory for gravitational accretion onto a moving star was described more than 70 years ago by \citet{hl39a}, 
from a theoretical standpoint we know surprisingly little about its stability and temporal behavior. 
The simplified case of planar accretion in two dimensions is known to be unstable to the growth of small perturbations from 
a steady state \citep{bp09}, leading to accretion through a quasi-Keplerian accretion disk.
Whether similar behavior exists in three dimensions remains an open question. 
Previous 3D simulations have observed rotational flow around the surface of the accretor, but were limited by 
the computational resources necessary for small accretor radii and sufficiently high resolutions. 
In this paper, we describe the results of high-resolution hydrodynamic simulations that probe the question: is three-dimensional 
Hoyle-Lyttleton accretion stable, or is the axisymmetry of the flow broken through rotational flow or violent flip-flop behavior?

A thorough review of Hoyle-Lyttleton accretion (HLA) theory and subsequent numerical studies is given by \citet{e04}. We provide only a brief synopsis here. 
Hoyle and Lyttleton derived a formula for the accretion radius of a point mass $M$ moving at a speed $V_\infty$ through a uniform medium of density $\rho_\infty$:

\begin{equation}
R_a=\frac{2GM}{V_\infty^2}.
\end{equation}

Gas approaching the star with an impact parameter less than $R_a$ will collide on an accretion line behind the star and will be left with insufficient kinetic energy to escape the gravitational potential of the star. From this, \citet{hl39a} posited that the mass accretion rate is given by the mass flux through a circle of radius $R_a$ far upstream:

\begin{equation}
\dot M_{HL} = \pi R_a^2 V_\infty \rho_\infty.
\label{eqn:mdot}
\end{equation}

There have been many numerical simulations of HLA in 2D, beginning with the axisymmetric steady-state solutions of \citet{h71}. 
\citet{mis87} were the first to observe what they called the 'flip-flop' instability, which was also seen by \citet{ft88}. 
Flip-flop behavior is characterized by an accretion wake that oscillates between alternately rotating disk states with a brief transitory phase in which the disk is flushed into the accretor. 
This non-axisymmetric instability has been subsequently seen in numerous hydrodynamic simulations of two-dimensional planar accretion \citep{mss91,zwn95,blt97,shima98,pom00}. 
\citet{bp09} showed that the flip-flop instability was a true overstability rather than a consequence of numerical effects, and that the growth rate increases as the radius of the accretor is decreased.

The relevance of instabilities in two-dimensional HLA to the temporal behavior of accreting X-ray sources has been debated since the discovery of the flip-flop instability. 
Given that accretion-powered X-ray sources in Be-type binary systems are expected to accrete from a planar disk expelled by the companion Be star, attention has focused on the relationship of the flip-flop instability to these Be-type X-ray pulsars.  
\citet{tfb88} interpreted the recurrent flares in EXO 2030+375 in terms of the episodic accretion resulting from the flip-flop instability.  
\citet{blt97} argued that the variable accretion torques associated with the flip-flop instability may explain the erratic spin behavior in these systems.

It remains to be seen whether this disk accretion mode exists in 3D.  
The best simulations to date of 3D HLA were a series of studies using nested cartesian grids \citep{ra94, r97, r99}.
The accretion in these 3D studies was found to be unstable, although not as violent as the flip-flop behavior found in 2D planar flow. 
Nonetheless, some of the 3D models reported in \citet[hereafter RA]{ra94} did exhibit a quasi-periodic swinging of the accretion shock 
leading to brief epochs of strong rotational flow near the accretor. This behavior was particularly evident for the models with the 
smallest accretors. The models with an accretor radius of $R_s = 0.05\, R_a$ (S6 and S8) exhibited rotational flow near the 
surface of the accretor, but the flow velocities were sub-Keplerian by more than a factor of two and the rotational flow pattern 
lasted for no more than 10 orbits.  

\defcitealias{ra94}{RA}

However, it is important to note that these simulations were of relatively low spatial resolution and limited to relatively large radii for the accreting star.  Although these 3D runs used nested cartesian grids to improve the resolution near the accreting star, each nested 
grid was only 32 zones on a side.  Their best resolution amounted to 6.4 zones across the radius of the accretor.  
This is very low compared to that used in many of the 2D simulations exhibiting flip-flop behavior (Blondin \& Pope 2009 used 
a resolution 6 times finer).   Moreover, these simulations were limited to an accretor radius of 0.05 $R_a$ (and one short run 
at 0.01 $R_a$) due to limited computational resources.  This limitation is important because the growth rate of the flip-flop 
instability in 2D increases dramatically with decreasing accretor radius \citep{bp09}, with long phases of quasi-disk accretion 
only occurring for much smaller accretors (0.0037 $R_a$).  

In this paper we describe two ($R_s/R_a$ = 0.05 and 0.01) high-resolution simulations of three-dimensional HLA.
The parameters of these models were chosen to mimic two of the simulations described by \citetalias{ra94} in which they
found time-dependent, non-axisymmetric behavior.

\section{Computational Method}

The simulations presented here use the VH-1 hydrodynamics code, which is the same code described in \citet{bp09}.
VH-1 uses a Lagrange-remap formulation of the piecewise parabolic method \citep{cw84} to solve the Euler equations for the 
inviscid flow of a compressible ideal gas. The piecewise parabolic method is an explicit Godunov method that offers high-order 
accuracy (3rd-order in space, 2nd-order in time). 

The extension to 3D is enabled by the use of a Yin-Yang overset grid as described by \citet{ks04}. 
Given that one needs to follow the accretion flow over orders of magnitude in distance from the central object, 
a spherical coordinate system is the natural choice for accretion studies.  Unfortunately in 3D this choice introduces a
coordinate singularity associated with the polar axis that creates severe problems with the Courant limit on the time step in 
explicit methods. It also results in unwanted artifacts associated with advection near the axis.  
The Yin-Yang approach solves these problems by using two spherical polar grids aligned by 90 degrees with respect to each other. 
The polar regions of one grid are replaced by the equatorial region of the other, 
with overlapping regions used to implement boundary conditions in the angular directions. 

We use Yin and Yang grids with dimensions of 252x72x216 zones, which when combined result in a grid circumference of 288 zones.
We use a non-uniform radial grid, keeping $\Delta r/r$ roughly constant throughout the grid.
This grid enables us to simulate the downstream behavior of the accretion shock extending to $10 R_s$ while still offering a resolution near the accretor comparable to that of the highest-resolution 2D simulations published to date.
The number of radial zones was chosen to produce a value of $\Delta r/r\approx 0.021$, comparable to the angular
spacing of $\pi/144$.  This ensures approximately cubic zones throughout the grid.
For the $R_s = 0.01\, R_a$ run we extended the radial grid to 324 zones to keep the same value of $\Delta r/r$. 

\begin{figure}[!htp]
\begin{center}
\includegraphics[width=84mm]{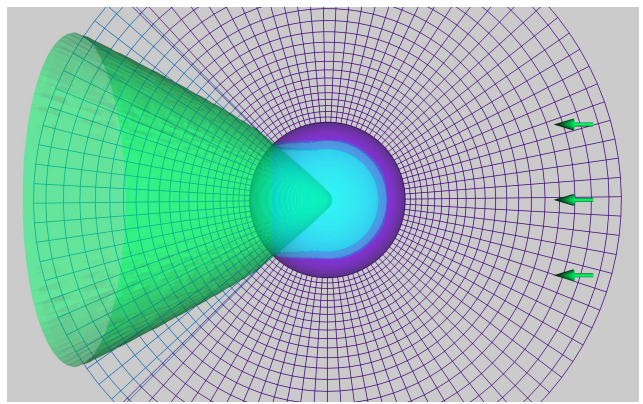}  
\caption{A schematic diagram of the grid orientation, as seen looking down the +z axis. The incoming flow begins on the Yin grid (purple) and flows toward the +x axis. The bow shock (green) is almost entirely contained within the Yang grid (blue). The grid extends to the outer edges of the highlighted plane. A portion of the grid has been made semi-opaque in the middle to illustrate the Yin-yang grid geometry.
(A color figure is available in the online journal.)}
\label{fig:schematic}
\end{center}
\end{figure}

We orient our grid so that the incoming gas flows toward the origin from the +x direction, such that the projected 
face of the Yin-Yang grid is laterally symmetric (Figure \ref{fig:schematic}). This places the accretion wake almost entirely in the Yang grid, and 
minimizes the impact of the grid seams on the subsonic post-shock flow. 
During preliminary runs, we observed the formation of small, low-entropy streamers that originated in the low-density stagnation 
region of the post-shock flow and propagated upstream toward the accretor. These streamers would occasionally wrap partly
around the accretor, disturbing the flow slightly as they were accreted and causing minor motion of the wake. 
The origin of these streamers appears to lie in the seams between the Yin and Yang grids, 
as reorienting the flow so that the seams are no longer present in the accretion wake causes them to vanish.

The inner boundary of our simulation represents the surface of the accretor.
In order to minimize the impact of this inner boundary on the accretion flow, we imposed an absorbing boundary condition
by setting the density and pressure to very small values inside of this boundary.
This is similar to the inner boundary condition used by \citet{h71}, \citetalias{ra94}, \citet{blt97}, and \citet{bp09}.
To allow gas to flow unimpeded off the grid downstream of the accretor, we implement a zero-gradient 
outer boundary downstream, but with the added condition that gas may not flow inward onto the grid once it has escaped. 

In order to monitor the accretion of mass and angular momentum, we record the total mass and angular momentum crossing 
the inner boundary over each time step. By averaging these values periodically over small intervals of time, we are able to obtain a measure of the mass and angular momentum accretion rates. 

To initialize the 3D grid, our approach is to begin with 2D axisymmetric simulations on a spherical grid.  These 2D simulations are evolved for several flow times to allow them to settle into a steady state (if such a state exists), after which the evolved 2D solution is mapped onto the 3D Yin-Yang grid. The 3D grid uses the exact same radial grid as in the 2D simulation, avoiding the need for interpolation in the radial direction.  Moreover, the 2D and 3D codes share the exact same FORTRAN code for hydrodynamic evolution. This procedure is meant to minimize any transients generated by the initial conditions on the 3D grid. Starting from a steady state allows us to investigate the origin of any instability by watching small perturbations grow over time.  It also allows us to avoid the relatively long initial evolution that previous researchers had to compute while waiting for large-amplitude transients to die away. 

Simulation parameters are chosen to match models S8 (S6 was the same model but lower resolution) and T10 of \citetalias{ra94}, with an upstream Mach number of three, a ratio of specific heats of 5/3, and accretor radii of $R_s = 0.05\,R_a$ and $ 0.01\,R_a$. 
Assuming a strictly uniform plane-parallel flow at the outer boundary has the adverse effect of coupling the behavior of the flow at the 
accretor boundary to the outer radius of the grid because the outer boundary is not truly an infinite distance away.  To more closely
match the true HLA problem, we calculate the velocity, density, and pressure (assuming isentropic flow) 
at the outer edge of the grid by solving for the ballistic trajectory from infinity \citep{bk79}. 

\section{Results}

\subsection{Large Accretor: $R_s/R_a = 0.05$}

\begin{figure}[!htp]
\begin{center}
\includegraphics[width=84mm]{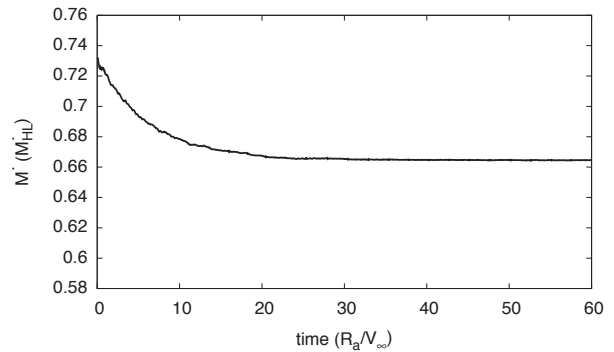}
\caption{The mass accretion rate as a function of time for the model with $R_s/R_a = 0.05$.}
\label{fig:largemdot}
\end{center}
\end{figure}

With the larger accretor ($R_s/R_a=0.05$, corresponding to S8 in \citetalias{ra94}), we observed a remarkably stable flow over 60 flow times. 
After interpolating the two-dimensional axisymmetric solution onto the 3D grid, the shock cone adjusts slightly over a few flow times. 
By a time of $40\, R_a/V_\infty$ the mass accretion rate has settled to a constant value of $0.67\, \dot M_{HL}$, as shown in Figure \ref{fig:largemdot}.   This is consistent
with the average value of $0.56\, \dot M_{HL}$ reported by \citetalias{ra94}, given the large variability seen in their models.  In contrast,
the mass accretion rate in our large accretor model varied by less than 0.02\% once it settled down to the steady-state value.
These small variations about the average could be attributed to numerical noise behind the steady bow shock.

\begin{figure}[!htp]
\begin{center}
\includegraphics[width=84mm]{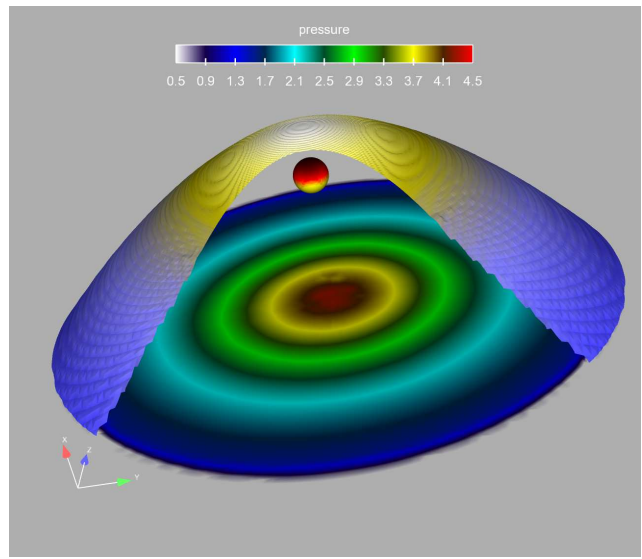}
\caption{The pressure in a 2D plane a distance $0.8\,R_a$ downstream of the accretor illustrates the highly axisymmetric 
nature of the downstream flow.  The incoming flow is moving down the x-axis.  A portion of the bow shock (on the Yin grid) is
is also shown.  The surface of the accreting star is shown with the leading face colored dark red to represent low mass flux and the back side
colored yellow to represent high (inward) mass flux.  
(A color figure is available in the online journal.)}
\label{fig:symmetry}
\end{center}
\end{figure}

This steady flow retains the original axisymmetry to a high degree.  The maximum specific angular momentum of the accreted gas never
exceeds 0.05\% of the specific angular momentum of a Keplerian orbit at the radius of the accretor.  This axisymmetry is illustrated in
Figure \ref{fig:symmetry}, which shows the gas pressure in a 2D plane normal to the incoming flow but downstream of the accretor.

\begin{figure*}
\begin{center}
\begin{minipage}{140mm}
\includegraphics[width=140mm]{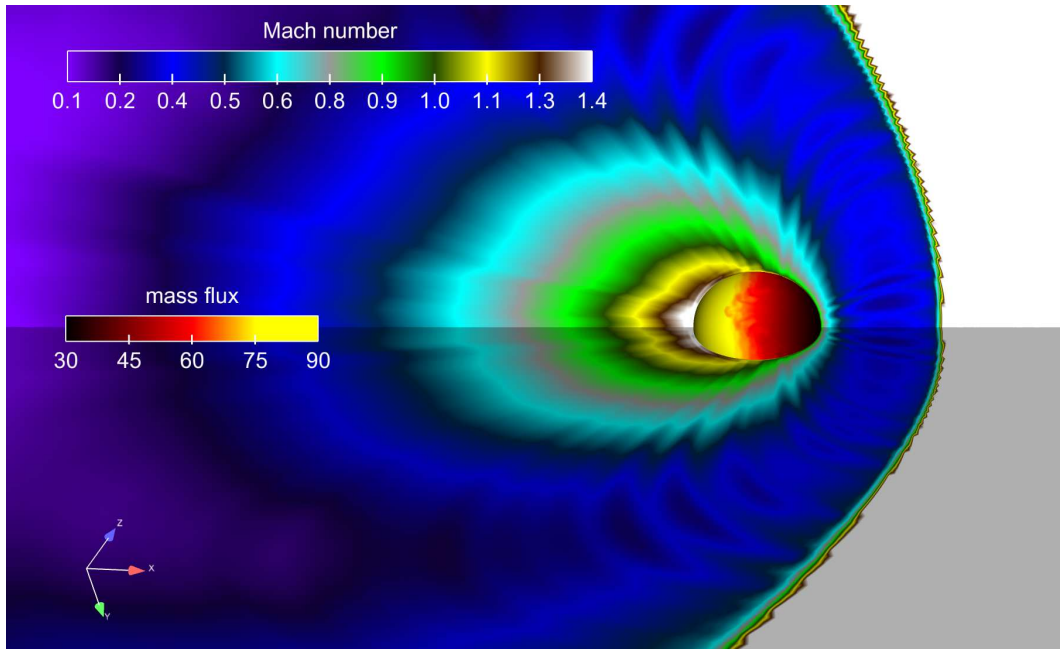}  
\caption{The Mach number in the post-shock flow for the large accretor, displayed on two orthogonal planes cutting through the flow axis. 
The surface of the accretor is colored by the local mass flux.  Most of the accretion occurs on the downstream side of 
the accretor, at Mach numbers slightly greater than unity.
(A color figure is available in the online journal.)}
\label{fig:machlarge}
\end{minipage}
\end{center}
\end{figure*}

The structure of this HLA flow is in reasonable agreement with time-independent, axisymmetric calculations \cite{h71}.
The local Mach number of the flow is shown in Figure \ref{fig:machlarge}, illustrating an upstream subsonic
stagnation region, and a downstream, supersonic accretion flow.  The region between the accretor surface and
the stand-off bowshock is highly subsonic.  The high pressure of this region results in a relatively low mass accretion
rate through the upstream half of the accretor surface.  In contrast, the inward accretion flow on the downstream side
of the accretor reaches a Mach number of $\sim 1.5$.  
Most of the mass accretion occurs preferentially on the downstream side of the accretor; the local mass flux on the
leading side is three times lower than that on the downstream side.
At equilibrium, the bow shock has a stand-off distance of $\sim 0.15\, R_a$.
While the opening angle of the bowshock is ill-defined in the vicinity of the accretor, at a distance of several $R_a$ downstream
the opening half-angle gradually decreases to a roughly constant $\sim 30^\circ$.
This is noticeably larger than the theoretical prediction of ${\rm sin^{-1}}(1/M)\approx 20^\circ$ for
a Mach number of $M=3$. (Petrich et. al. 1988 also noted a larger opening angle).

\subsection{Small Accretor: $R_s/R_a = 0.01$}

Our second simulation reduced the accretor size to a value of $R_s/R_a = 0.01$, the same as in the T10 simulation of RA. At the time, RA's simulation lacked the duration to draw appropriate conclusions, although it displayed behavior that supported their hypothesized trend of more volatile behavior in $\dot M$ and $\dot J$ with decreasing accretor size. 

\begin{figure*}
\begin{center}
\begin{minipage}{140mm}
\includegraphics[width=140mm]{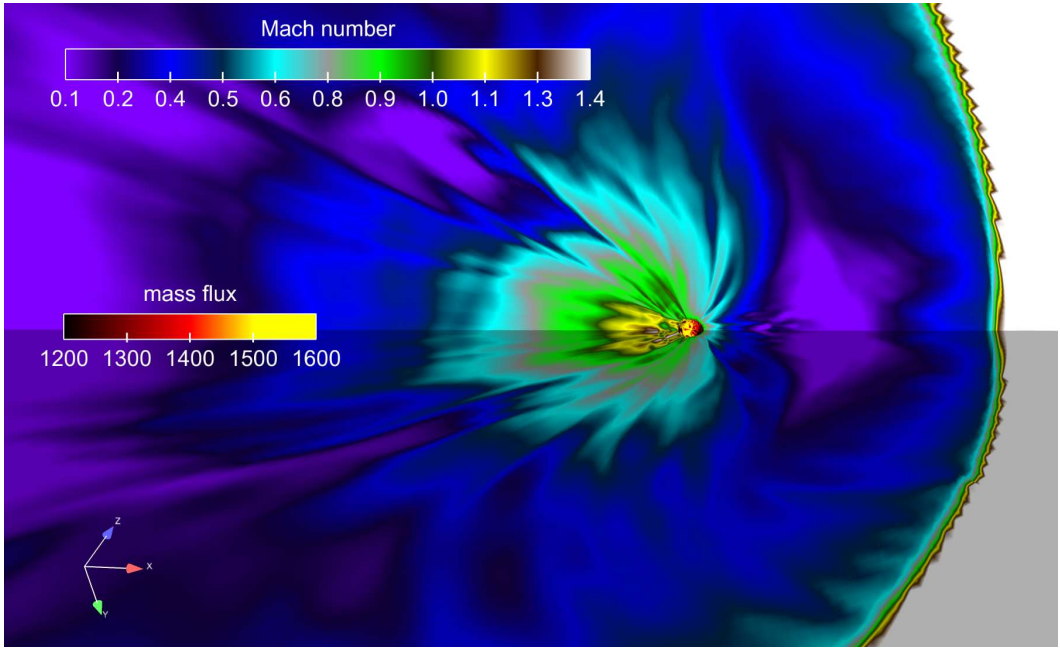}  
\caption{The Mach number in the post-shock flow for the small accretor, displayed on two orthogonal planes cutting through the flow axis. 
The surface of the accretor is colored by the local mass flux.
(A color figure is available in the online journal.)}
\label{fig:machsmall}
\end{minipage}
\end{center}
\end{figure*}

Our simulation extends to 17 flow times, roughly three times longer than in \citetalias{ra94} and long enough to see
the long term behavior unrelated to the initial conditions.  
The general structure of the accretion flow, illustrated in Figure \ref{fig:machsmall}, remains similar to that seen in the larger accretor.
The shock front is located at an upstream distance of $\sim 0.25\, R_a$, significantly farther out than in
the large-accretor simulation. The downstream opening half-angle is $\sim 34^\circ$.  Again the mass accretion occurs preferentially on the
downstream side, with Mach numbers approaching 1.5.  

\begin{figure}[!htp]
\begin{center}
\includegraphics[width=84mm]{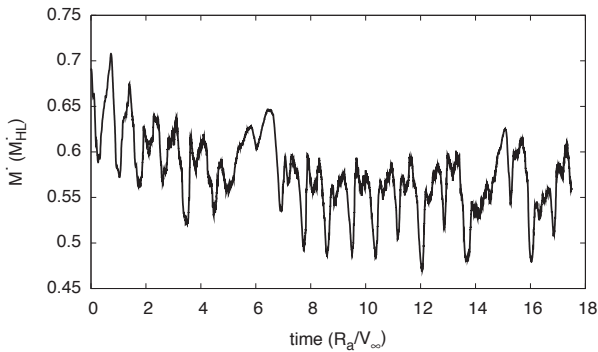}  
\caption{The mass accretion rate as a function of time for the model $R_s/R_a = 0.01$.}
\label{fig:smallmdot}
\end{center}
\end{figure}

The most noticeable difference in this flow is the presence of a breathing mode, 
in which the accretion wake oscillates in the direction of the original flow while remaining axisymmetric. 
This oscillation begins immediately, and while it persists throughout the simulation, its amplitude remains finite and the large-scale structure of the wake remains undisrupted.   The stand-off distance of the bow shock varies quasi-periodically with an amplitude of about 10\% of the
steady-state value and an oscillation period slightly smaller than $R_a/V_\infty$.
The streaks in Figure \ref{fig:machsmall} betray the unsteady nature of the accretion flow
onto the small accretor in contrast to the very steady flow onto the large accretor.

The mass accretion rate, plotted in Figure \ref{fig:smallmdot}, follows the same general evolution as for the large accretor.  The 
accretion rate initially decreases until settling into a constant average after about ten flow times.  In this case the late-time average is
about $0.56\, \dot M_{HL}$, somewhat lower than for the larger accretor.  The main difference is an oscillation
imposed on top of this long term behavior.  These oscillations in the mass accretion rate have the same period as the oscillations
in the shock stand-off distance, hence we attribute both to the axisymmetric breathing mode.  The phase difference between
mass accretion and shock position is such that  the brief minima in the mass accretion rate corresponds to when the leading
edge of the bowshock is falling back in toward the accretor.

Although the flow is no longer steady, it remains highly axisymmetric.
The accreted angular momentum is about an order of magnitude larger than in the large accretor simulations, but the values
remain relatively small and there is no preferred orientation of the rotational flow.  
Even with the smaller radius of accretor surface, the instantaneous value of the specific angular momentum of accreting gas 
remains less than 5\% of the specific angular momentum of a Keplerian orbit at the accretor surface.

\section{Conclusions}

We present high-resolution simulations of Hoyle-Lyttleton Accretion that demonstrate that the axisymmetry of HLA is
remarkably stable. 
For both simulations we observe no rotational flow in the vicinity of the accretor, 
as indicated by the small amount of specific angular momentum in the accreted gas. 
For the large accretor case ($R_s=0.05R_a$) the mass accretion rate relaxes to a constant value of $0.67 \dot{M}_{HL}$, 
with variations no greater than 0.02\%.  

Decreasing the accretor size to $R_s=0.01R_a$ introduces a quasi-periodic axisymmetric breathing mode. This mode appears 
almost immediately and reaches a quasi-stable oscillation within 10 $R_a/V_\infty$. The mass accretion rate is modulated by the 
oscillation of the bow shock, but maintains an average of $0.56 \dot{M}_{HL}$. The oscillations occur with a period of $0.86 R_a/v_\infty$. 
This is decidedly larger than the acoustic period (the time for a sound wave to travel from the bowshock to the accretor and back), which we 
calculate to be $0.26 R_a/V_\infty$. This breathing mode does not affect axisymmetry of the flow.

Our results are dramatically different from those of RA, who used a similar numerical algorithm but a different grid 
geometry (nested cartesian grids versus overset spherical grids) with substantially lower spatial resolution. 
Their simulations with $R_a/R_s=0.05$ and $0.01$ both displayed quiescent periods broken by brief transient disk-like flow. 
The larger accretor simulation in particular displayed an anti-correlation between the mass accretion rate and the specific angular 
momentum of the accreted gas: during the active periods when the accreting gas had a larger specific angular momentum 
there was a drop (by a factor of 2 or 3) in the mass accretion rate.  This behavior was not as dramatic as the flip-flop flow
observed in 2D HLA simulations.  In contrast, our simulations do not display rotational flow and remain almost entirely axisymmetric 
throughout their runtimes.

The simulations of RA included a random density perturbation of 3\% in their initial upstream flow as a means of 
breaking the initial symmetry of the flow. Our simulations do not include any initial noise, 
but some intrinsic noise is generated at the bow shock due to the numerical algorithm. The effects of this noise can be seen 
in Figure \ref{fig:machsmall}, in which the asymmetry is visible near the downstream surface of the accretor. 
The addition of small density perturbations is likely to be insufficient to significantly affect the stability of the flow, 
and would likely have an effect similar to that of the numerical noise. The simulations of \citet{r97,r99} suggest  that 
large-scale density and velocity gradients (3\% and 20\% across the upstream accretion column) may be more likely to 
instigate a non-stationary flow. These simulations displayed a relatively stable bow shock and no flip-flop behavior. 

The presence of the breathing mode and slight asymmetries near the accretor in the small-accretor case 
indicate that our results follow the general trend present in 2D simulations: a smaller accretor produces a less stable flow pattern. 
This supports the notion that extending 3D simulations to smaller accretor 
radii is critical for understanding the true behavior of HLA in astrophysical contexts. 
The continued investigation of its stability under the influence of greater perturbations (including large scale gradients) 
and smaller accretor size will allow us to reassess what conditions, if any, produce an unstable flow and possible accretion of angular momentum. In addition, introducing strong localized density perturbations analogous to those of a clumpy stellar wind may provide insight into the nature of supergiant fast x-ray transient outbursts while simultaneously addressing fundamental questions regarding stability.

\acknowledgments

This work was supported in part by NSF grant OCI-749248.   
This research was supported by an allocation of advanced computing resources provided by the National Science Foundation. 
The computations were performed on Kraken at the National Institute for Computational Sciences.

\clearpage

\end{document}